\documentclass[twocolumn,superscriptaddress,aps,floatfix,preprintnumbers,amsmath,amssymb,prl,10pt]{revtex4-1}
\usepackage[titletoc,toc,title]{appendix}
\usepackage[english]{babel}
\usepackage[hidelinks]{hyperref}
\usepackage[normalem]{ulem}
\usepackage{adjustbox}
\usepackage{amsfonts,amsmath,amssymb}
\newcommand{\msun}{M_{\odot}}
\usepackage{bm}
\usepackage{enumitem}
\usepackage{epsfig}
\usepackage{cancel}
\usepackage{centernot}
\usepackage{color}
\usepackage{comment}
\usepackage{contour}
\usepackage{flushend}
\usepackage{footmisc}
\usepackage{graphics}
\usepackage{graphicx}
\usepackage{mathrsfs}
\usepackage{mdframed}
\usepackage[normalem]{ulem}
\usepackage{pifont}
\usepackage{relsize}
\usepackage{shadowtext}
\usepackage{slashed}
\usepackage{soul}
\usepackage{subfigure}
\usepackage{tcolorbox}
\usepackage{textgreek}
\usepackage{titlesec}
\usepackage{titletoc}
\usepackage{verbatim}
\usepackage{xcolor}
\usepackage{xfrac}
\contourlength{0.2em}
\usepackage{orcidlink}
\usepackage{ragged2e}

\newtcolorbox{mybox}[3][]{
  colframe = black,
  colback  = #2!10,
  coltitle = #2!20!black,
  title    = {#3},
  boxsep   = 2pt,
  top      = 1pt,
  bottom   = 1pt,
  boxrule  = 0.8pt,  
  #1,
}

\definecolor{zima_blue}{HTML}{1393C1}
\hypersetup{setpagesize=false,bookmarksnumbered=true,bookmarksopen=true,colorlinks=true,linkcolor=zima_blue,urlcolor=zima_blue,citecolor=zima_blue,linktocpage=false}

\newif\ifarxiv
\arxivfalse     

\begin{document}

\title{Closing the Mass Window for Stupendously Large Black Holes}

\author{Christopher Gerlach\orcidlink{0009-0000-3431-2174}}
\email{cgerlach@uni-mainz.de} 
\affiliation{PRISMA$^+$ Cluster of Excellence \& Mainz Institute for Theoretical Physics, Johannes Gutenberg-Universität Mainz, 55099 Mainz, Germany}
\author{Yann Gouttenoire\orcidlink{0000-0003-2225-6704}}
\email{yann.gouttenoire@gmail.com}
\affiliation{PRISMA$^+$ Cluster of Excellence \& Mainz Institute for Theoretical Physics, Johannes Gutenberg-Universität Mainz, 55099 Mainz, Germany}
\author{Antonio J. Iovino\orcidlink{0000-0002-8531-5962}}
\email{ai2869@nyu.edu} 
\affiliation{New York University, Abu Dhabi, PO Box 129188 Saadiyat Island, Abu Dhabi, UAE}
\affiliation{Department of Theoretical Physics and Gravitational Wave Science Center,  \\
24 quai E. Ansermet, CH-1211 Geneva 4, Switzerland}

\author{Nicholas Leister\orcidlink{0009-0007-6113-313X}}
\email{nleister@uni-mainz.de} 
\affiliation{PRISMA$^+$ Cluster of Excellence \& Mainz Institute for Theoretical Physics, Johannes Gutenberg-Universität Mainz, 55099 Mainz, Germany}

\begin{abstract}
We show that primordial black holes (PBHs) in the \emph{Stupendously Large Black Hole} mass range ($M \gtrsim 10^{11}\,M_\odot$) produce isocurvature perturbations exceeding current \textit{Planck} Cosmic Microwave Background (CMB) limits, thereby excluding them as a significant dark matter component.

\vspace*{5pt}
\noindent
\textit{Published version:~\href{https://doi.org/10.1103/qlf9-thf4}{Phys. Rev. Lett. 137, 181401}}

\end{abstract}

\preprint{MITP-25-053}

\maketitle

\noindent{{\bf{\em Introduction.}}}
\label{sec:Introduction} 
PBHs~\cite{Zeldovich:1967lct,Hawking:1974rv,Chapline:1975ojl,Carr:1975qj,Carr:2026hot} have attracted significant attention in recent years, both as potential contributors to the dark matter (DM) content of the universe and as possible explanations for some of the black hole merger events observed by LIGO/Virgo/KAGRA~\cite{Bird:2016dcv, Sasaki:2016jop, Clesse:2016vqa,Raidal:2018bbj,Vaskonen:2019jpv,Hutsi:2020sol,Mukherjee:2021ags,Pujolas:2021yaw,Romero-Rodriguez:2021aws,DeLuca:2020sae,DeLuca:2020agl,Franciolini:2021tla}.

A wide range of astrophysical and cosmological observations tightly constrain the present-day PBH dark matter fraction. For non-spinning PBHs, the only mass range in which they could constitute all of the DM is the \emph{asteroidal window}, spanning $10^{-16}$–$10^{-12}\,\msun$.

Another mass range that has recently attracted attention is that of \emph{stupendously large black holes} (SLABs), defined by $M_{\rm PBH} \gtrsim 10^{11}\,\msun$~\cite{Carr:2020erq}. Being too massive to reside within galaxies, they cannot constitute the totality of the DM but, according to Ref.~\cite{Carr:2020erq}, could still comprise a significant DM fraction for $M_{\rm PBH}\in[10^{14},10^{19}]\,\msun$~\footnote{ Ref.~\cite{Carr:2020erq} also considered accretion-based bounds, but these are highly uncertain due to the limitations of the Bondi formalism in an expanding Universe, radiation drag, and uncertainties in accretion efficiency and spectral modeling~\cite{Serpico:2020ehh,DeLuca:2020fpg,Agius:2024ecw}.}.

A key property of PBHs is that they generate \emph{isocurvature} perturbations: local overdensities in PBHs are initially compensated by radiation deficits, yielding no net curvature perturbation on large scales~\cite{Afshordi:2003zb}. As the Universe expands and radiation dilutes, this compensation is lifted, and PBH fluctuations source gravitational potentials that imprint characteristic signatures on the CMB.

In this \textit{Letter}, we show that isocurvature perturbations from SLABs exceed current limits from the CMB and Baryonic Acoustic Oscillation (BAO), thereby ruling them out as a significant DM component.

\noindent{{\bf{\em PBH Isocurvature Perturbations.}}} 
We model the total cold DM as a mixture of PBHs and particle dark matter (PDM, such as WIMPs or ALPs), 
\begin{equation}
    \rho_{\rm DM} = \rho_{\rm PBH} + \rho_{\rm PDM}, \quad
    f_{\rm PBH} \equiv \frac{\rho_{\rm PBH}}{\rho_{\rm DM}},
\end{equation}
such that the total DM density contrast is
\begin{equation}
\label{eq:delta_DM_def}
    \delta_{\rm DM} = f_{\rm PBH}\,\delta_{\rm PBH} + (1-f_{\rm PBH})\,\delta_{\rm PDM}.
\end{equation}
Following the linear perturbation theory for a PBH–PDM mixture presented in~\cite{Inman:2019wvr}, we assume that the primordial PDM perturbations are purely adiabatic, $\delta_{\rm PDM}^f=\delta_{\rm ad}^f$, where the superscript $f$ denotes the epoch of PBH formation.
The discreteness of PBHs induces an additional isocurvature component in their primordial perturbations~\cite{Tada:2015noa,Young:2015kda}~\footnote{The presence of non-local non-Gaussianities in the statistics of the large fluctuations that produce PBHs can also source isocurvature perturbations~\cite{Tada:2015noa,Young:2015kda}.}
\begin{equation}
    \delta_{\rm PBH}^f = \delta_{\rm ad}^f + \delta_{\rm iso}^f,
\end{equation}
so that Eq.~\eqref{eq:delta_DM_def} provides the primordial DM contrast with an isocurvature component as well
\begin{equation}
    \delta_{\rm DM}^f = \delta_{\rm ad}^f + f_{\rm PBH}\,\delta_{\rm iso}^f.
\end{equation}
Introducing the linear transfer functions for adiabatic and DM-isocurvature modes between epoch of PBH formation $a_f$ and a later epoch $a$, $T_{\rm ad}(a)$ and $T_{\rm iso}(a)$, the DM contrast evolves as
\begin{equation}
\label{eq:delta_DM_evol}
    \delta_{\rm DM}(a) = T_{\rm ad}(a)\,\delta_{\rm ad}^f
    + T_{\rm iso}(a)\,f_{\rm PBH}\,\delta_{\rm iso}^f.
\end{equation}
The DM isocurvature perturbation is defined as the fluctuation in the DM to photon number density ratio $n_{\rm DM}/n_\gamma$~\cite{Planck:2018jri},
\begin{equation}
\label{eq:IDM_def}
    \mathcal{I}_{\rm DM} \equiv \frac{\delta(n_{\rm DM}/n_\gamma)}{n_{\rm DM}/n_\gamma}
    = \delta_{\rm DM} - \frac{3}{4}\,\delta_{\rm R}.
\end{equation}
Substituting Eq.~\eqref{eq:delta_DM_evol} into Eq.~\eqref{eq:IDM_def}, using that the adiabatic contribution cancels, yields
\begin{equation}
\label{eq:IDM_a}
    \mathcal{I}_{\rm DM}(a) = T_{\rm iso}(a)\,f_{\rm PBH}\,\delta_{\rm iso}^f.
\end{equation}
We define the dimensionless cross--power spectrum of two observables $X$ and $Y$ as  
\begin{equation}
\label{eq:P_II_def}
\mathcal{P}_{XY}(k) \equiv \frac{k^3}{2\pi^2} \int d^3\mathbf{r} \, e^{i\mathbf{k} \cdot \mathbf{r}} \left< X(0)\, Y(\mathbf{r})\right>,
\end{equation}
such that $\mathcal{P}_{XY}(k)\, d\ln k$ gives the contribution to the variance per logarithmic interval in $k$. For brevity, we denote $\langle XY \rangle \equiv \mathcal{P}_{XY}(k)$ when no ambiguity arises.

For simplicity, we model PBHs as point-like objects (the finite-size case is discussed in the Supplemental Material). The isocurvature density contrast at formation is then
\begin{equation}
\delta_{\rm iso}^f(\mathbf{x}) = \frac{1}{n_{\rm PBH}}\sum_i \delta^{(3)}(\mathbf{x}-\mathbf{x}_i) - 1,
\end{equation}
where $n_{\rm PBH}$ is the PBH number density, ensuring $\langle \delta_{\rm iso}^f \rangle = 0$.
Assuming a Poisson distribution, the PBH number-density field can be written as
\begin{equation}
n(\mathbf{x}) \equiv \sum_i \delta^{(3)}(\mathbf{x}-\mathbf{x}_i),
\qquad
\langle n(\mathbf{x}) \rangle = n_{\rm PBH}.
\end{equation}
Its two-point function reads
\begin{equation}
\left\langle n(\mathbf{x}_1)n(\mathbf{x}_2) \right\rangle
=
\left\langle \sum_{i,j}
\delta^{(3)}(\mathbf{x}_1-\mathbf{x}_i)\,
\delta^{(3)}(\mathbf{x}_2-\mathbf{x}_j)
\right\rangle.
\end{equation}
We separate the sum into diagonal ($i=j$) and off-diagonal ($i\neq j$) contributions.
For the diagonal part, using $\delta^{(3)}(\mathbf{x}_1-\mathbf{x}_i)\delta^{(3)}(\mathbf{x}_2-\mathbf{x}_i)
= \delta^{(3)}(\mathbf{x}_1-\mathbf{x}_2)\,\delta^{(3)}(\mathbf{x}_1-\mathbf{x}_i)$, we obtain
\begin{equation}
\left\langle \sum_i
\delta^{(3)}(\mathbf{x}_1-\mathbf{x}_i)
\delta^{(3)}(\mathbf{x}_2-\mathbf{x}_i)
\right\rangle
=
n_{\rm PBH}\,\delta^{(3)}(\mathbf{x}_1-\mathbf{x}_2).
\end{equation}
For the off-diagonal part, statistical independence implies factorization,
\begin{equation}
\left\langle \sum_{i\neq j}
\delta^{(3)}(\mathbf{x}_1-\mathbf{x}_i)
\delta^{(3)}(\mathbf{x}_2-\mathbf{x}_j)
\right\rangle
=
n_{\rm PBH}^2.
\end{equation}
Combining both contributions, we obtain
\begin{equation}
\left\langle n(\mathbf{x}_1)n(\mathbf{x}_2) \right\rangle
=
n_{\rm PBH}\,\delta^{(3)}(\mathbf{x}_1-\mathbf{x}_2)
+
n_{\rm PBH}^2.
\end{equation}Applying the definition of $\delta_{\rm iso}^f$, this directly yields
\begin{equation}
\left\langle \delta_{\rm iso}^f(0)\delta_{\rm iso}^f(\mathbf{r}) \right\rangle
=
\frac{1}{n_{\rm PBH}}\,\delta^{(3)}(\mathbf{r}).
\end{equation}
Inserting this result into Eq.~\eqref{eq:P_II_def}, we obtain
\begin{equation}
\left<(\delta_{\rm iso}^{f})^2\right>
=
\frac{k^3}{2\pi^2 n_{\rm PBH}}
=
\frac{2}{3\pi}\left( \frac{k}{k_{\rm PBH}}\right)^3,
\end{equation}
corresponding to a white-noise spectrum. We have introduced the characteristic interspacing scale
\begin{equation}
\label{eq:kPBB_main}
k_{\rm PBH} \equiv \left(\frac{4\pi }{3}n_{\rm PBH}\right)^{1/3} = \left(\frac{4\pi f_{\rm PBH} \,\rho_{\rm DM,0}}{3M_{\rm PBH}}\right)^{1/3},
\end{equation}
with $\rho_{\rm DM,0}\simeq 3\times10^{10} \,\msun\, \textrm{Mpc}^{-3}$ denoting the present-day mean dark matter density\,\cite{ParticleDataGroup:2020ssz}.
Hence, the DM isocurvature power spectrum in Eq.~\eqref{eq:IDM_a} becomes
\begin{equation}
\label{eq:IDM_final_result2}
\left<\mathcal{I}_{\rm DM}^2\right>
\simeq
T_{\rm iso}^2(a)\,f_{\rm PBH}^2\,
\left<(\delta_{\rm iso}^{f})^2\right>
\Theta(k_{\rm NL}-k)\,
\Theta\!\left(k_S(a)-k\right).
\end{equation}
We introduce two cutoffs. The first, $k_{\rm NL}$, excludes modes entering the non-linear (NL) regime. The second, $k_S(a)\equiv a\,R_S^{-1}$ (normalized to $a=1$ today), corresponds to the inverse Schwarzschild radius and accounts for PBH exclusion effects. For a conservative estimate, we evaluate the latter at $a=a_f$ (see Eq.~\ref{eq:kf} in the Supplemental Material)~\footnote{We find that this second cutoff does not affect our constraints as $k_{\rm NL}<k_S(a_f)$ for both case $A$ and $B$ in all the relevant parameter space.}.
To evaluate $k_{\rm NL}$, we adopt two cutoff prescriptions, defined as the scales at which either the PBH (case A) or the DM (case B) fluctuations become nonlinear, i.e. exceed unity. 
\\ \textit{i)} For \textit{case A}, the NL cut-off is defined by 
\begin{align}
\label{eq:NL_PBH_def}
        \left<\delta_{\rm PBH}^2\right>(k_{\rm NL}^{\rm PBH}; a_{\rm rec}) \equiv 1,
\end{align}
where $a_{\rm rec}$ is the scale factor at recombination.
We now derive the evolution of $\delta_{\rm PBH}(a)$ following~\cite{Inman:2019wvr}. Assuming a negligible relative velocity between PDM and PBHs, the two fluids are subject to the same dynamics, leading to the conservation law
\begin{equation}
\label{eq:delta_minus_evol}
    \delta_{\rm PBH}(a)-\delta_{\rm PDM}(a) =\delta_{\rm PBH}^f-\delta_{\rm PDM}^f= \delta_{\rm iso}^f.
\end{equation}
From Eqs.~\eqref{eq:delta_DM_evol} and \eqref{eq:delta_minus_evol}, we get
\begin{align}
\label{eq:delta_PDM_a}
 \delta_{\rm PDM}(a) &=  T_{\rm ad}(a) \delta_{\rm ad}^f + (T_{\rm iso}(a)-1) f_{\rm PBH}\delta_{\rm iso}^f,\\
 \label{eq:delta_PBH_a}
    \delta_{\rm PBH}(a) &=  \delta_{\rm iso}^f+\delta_{\rm PDM}(a).
\end{align}
Plugging Eq.~\eqref{eq:delta_PBH_a} into Eq.~\eqref{eq:NL_PBH_def}, neglecting the adiabatic piece, at  leading order in $\mathcal{O}(f_{\rm PBH})$, we obtain
\begin{equation}
    ({A})\qquad k_{\rm NL}^{\rm PBH} \simeq \left(\frac{3\pi}{2}\right)^{1/3}k_{\rm PBH},
    \label{eq:cutoffA}
\end{equation}
where $k_{\rm PBH}$ is defined in Eq.~\eqref{eq:kPBB_main}.  This choice, however, might be too conservative since from solving the equation of motion of particles around PBHs,  Ref.~\cite{Inman:2019wvr}  finds that the isocurvature component of $\delta_{\rm PDM}(a)$ in Eq.~\eqref{eq:delta_PDM_a} evolves identically as in the linear regime for $k>k_{\rm NL}^{\rm PBH}$. 
\\ \textit{ii)} In \textit{case B}, we consider the breakdown of the linearity of the DM fluid
\begin{align}
\label{eq:NL_DM_def}
   \left<\delta_{\rm DM}^2\right>(k_{\rm NL}^{\rm DM}; a_{\rm rec}) \equiv 1.
\end{align}
{This aligns with previous definitions of the non-linear scale~\cite{ DeLuca:2020jug,Passaglia:2021jla,Domenech:2021and}.}
Plugging Eq.~\eqref{eq:delta_DM_evol} into Eq.~\eqref{eq:NL_DM_def} and ignoring the adiabatic piece~\footnote{We refer to Ref.~\cite{Liu:2022okz} for the study of the mixing between adiabatic and isocurvature components.}, we obtain  at leading order in $f_{\rm PBH}$
\begin{equation}
\label{eq:k_NL_intermediate}
   k_{\rm NL}^{\rm DM} =\frac{k_{\rm NL}^{\rm PBH}}{\left(f_{\rm PBH}T_{\rm iso}(a_{\rm rec})  \right)^{2/3}}.
\end{equation}
Neglecting effects from the baryons for simplicity, the isocurvature linear transfer function between the epoch of PBH formation $a_f$ and a given scale factor $a$ is~\cite{Meszaros:1974tb,Inman:2019wvr}
\begin{equation}
\label{eq:T_iso}
T_{\rm iso}(a) = \frac{2+3y}{2+3y_f},
\end{equation}
with $y_f\equiv a_f/a_{\rm eq}$ and $y\equiv a/a_{\rm eq}$. Plugging Eq.~\eqref{eq:T_iso} into Eq.~\eqref{eq:k_NL_intermediate}, we deduce
\begin{equation}
 (B)\qquad  k_{\rm NL}^{\rm DM} = k_{\rm NL}^{\rm PBH} \left(f_{\rm PBH}^{-1}\frac{2+3y_{\rm f}}{2+3y_{\rm rec}}  \right)^{2/3},
 \label{eq:cutoffB}
\end{equation}
where 
\begin{equation}
\label{eq:yf_def}
    y_{\rm f} = \gamma_{\rm H}^{-1/2}\!\left(\!\frac{g_{\star}(T_f)}{3.38}\!\right)^{\!\frac{1}{4}}\!\!\left(\!\frac{3.94}{g_{\star,s}(T_f)}\!\right)^{\!\frac{1}{3}}\!\!\!\left(\!\frac{M_{\rm PBH}}{4.1\times 10^{18}~M_{\odot}}\!\right)^{\!\frac{1}{2}}\!,
\end{equation}
with $g_{\star,s} (k)$ being the number of entropic degrees of freedom at PBH formation. 
The constraints in Ref.~\cite{Carr:2018rid} distinguish between structure formation driven collectively by Poisson fluctuations or individually via the ``seed effect'', depending on whether $k_{\rm NL}^{\rm DM} < k_{\rm NL}^{\rm PBH}$ or $k_{\rm NL}^{\rm DM} > k_{\rm NL}^{\rm PBH}$. These bounds require PBHs to trigger nonlinear DM structure formation, $k \gg k_{\rm NL}^{\rm DM}$. In contrast, our CMB limits are obtained in the linear DM regime ($k \ll k_{\rm NL}^{\rm DM}$), where this distinction is irrelevant and no seed effect operates.
 We checked that higher order corrections in $f_{\rm PBH}$ only slightly impact the cut-offs $k_{\rm NL}^{\rm PBH}$ and $k_{\rm NL}^{\rm DM}$ in Eqs.~\eqref{eq:cutoffA} and \eqref{eq:cutoffB} and do not impact our constraints.

\begin{figure}[t!]
\centering
\includegraphics[width=0.49\textwidth]{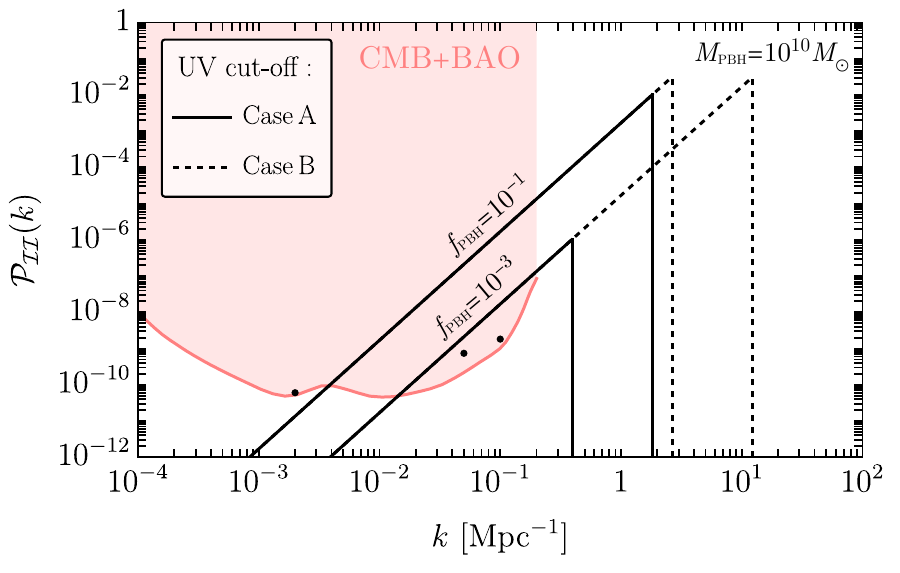}
\caption{\it CMB+BAO $95\%$ CL constraints on primordial DM isocurvature power spectrum taken from Ref.~\cite{Buckley:2025zgh}. For comparison, the three black dots show the constraints derived by the Planck collaboration~\cite{Planck:2018jri}. We show with black lines two representative cases for the PBH-induced isocurvature power spectrum defined in Eq.\,\eqref{eq:IDM_final_result2} with the case A cut-off (solid), Eq.\,\eqref{eq:cutoffA}, and case B cut-off (dashed), Eq.\,\eqref{eq:cutoffB}, for $M_{\rm PBH}=10^{10}\msun$ with $f_{\rm PBH}=10^{-1}$ and $f_{\rm PBH}=10^{-3}$.}\label{Fig:power}
\end{figure}
\begin{figure*}[t!]
\centering
\includegraphics[width=0.95\textwidth]{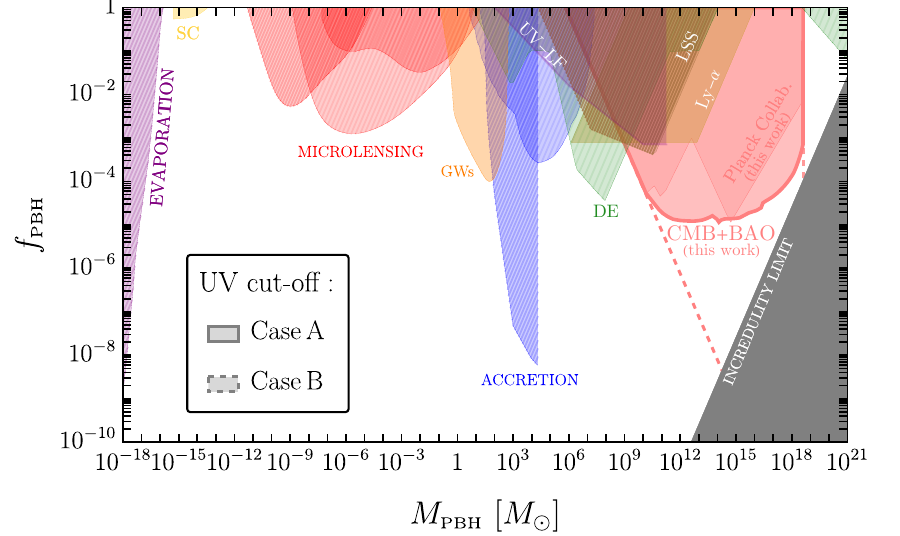}
\caption{\it  Current constraints on the abundance of a monochromatic, non-spinning PBH population. The new bounds from PBH-induced isocurvature perturbations derived in this work are shown in pink ({\bf CMB+BAO}). The thicker curves use the broad-$k$ constraints of Fig.~\ref{Fig:power} for the conservative (case~A, solid) and stringent (case~B, dashed) non-linear cut-offs. The thinner curve labeled ``Planck Collab.'' shows the bounds obtained from the \textit{Planck} three-point constraints in Eq.~\eqref{eq:beta_iso_constraints}, assuming case~A.
 Other constraints include bounds from PBH evaporation (magenta), PBH capture by stars (yellow), microlensing searches (red), GW signals from PBH mergers (orange), baryonic accretion (blue), dynamical effects (green), Hubble Space Telescope Ultraviolet
Luminosity Function (purple), large scale structure bounds (brown), and Lyman-$\alpha$ forests (dark gold). The grey region represents  the incredulity limit, which corresponds to having less than one single PBH within the current Hubble volume.}
\label{Fig:constraints}
\end{figure*}

\noindent{{\bf{\em New Isocurvature Constraints.}}}
Isocurvature perturbations are constrained by CMB temperature anisotropy, polarization, and gravitational lensing measurements from the \textit{Planck} satellite~\cite{Planck:2018jri}. We use the CMB/BAO constraints on the primordial isocurvature power spectrum $\mathcal{P}_{\mathcal{I}\mathcal{I}}(k)$ from Ref.~\cite{Buckley:2025zgh}, obtained over a broad range of $k$ using CLASS \cite{Blas:2011rf} and MontePython~\cite{Brinckmann:2018cvx} for MCMC fits to CMB~\cite{Planck:2018vyg} and BAO~\cite{BOSS:2016wmc} data. 

The bounds in Ref.~\cite{Buckley:2025zgh} assume isocurvature modes initialized deep in the radiation era. In our setup, they are generated at PBH formation ($y_f$), leading to a reduced amplitude at the time of recombination ($y_{\rm rec}$). This is accounted for by suppressing the DM isocurvature power spectrum in Eq.~\eqref{eq:IDM_final_result2} with a factor $\mathcal{S}(y_f)\le 1$,
\begin{equation}
    \label{eq:supp_S_Planck_main}
    \left<\mathcal{I}_{\rm DM}^2\right>~\rightarrow~
    \mathcal{S}^2(y_f)\,\left<\mathcal{I}_{\rm DM}^2\right>.
\end{equation}
Following~\cite{Peter:2013avv}, we calculate the factor $\mathcal{S}(y_f)$ by numerically solving the coupled evolution of the metric perturbation $\Phi$ and the isocurvature mode $\mathcal{I}_{\rm DM}$ from an early radiation-dominated epoch up to recombination. The numerical result is accurately captured by the simple fit (see Eq.~\eqref{eq:supp_S_ana} in the Supplemental Material for details)
\begin{equation}
    \label{eq:supp_S_fit_main}
    \mathcal{S}(y_f)
    \simeq
    \big(1-(y_f/y_{\rm rec})^{3/4}\big)^2,
    \qquad y_{\rm rec}\simeq 3.1.
\end{equation}
Isocurvature is modeled as a Dirac function, $\mathcal{P}_{\mathcal{I}\mathcal{I}}^{\rm iso}(k)=\mathcal{A}_{\rm iso}\,\delta(\ln k-\ln k_0)$, and $95\%$~CL constraints on $(\mathcal{A}_{\rm iso},k_0)$ are shown in Fig.~\ref{Fig:power}. Since any spectrum can be decomposed as the linear combination $\mathcal{P}_{\mathcal{I}\mathcal{I}}^{\rm iso}(k)=\int d\ln k_0\,\mathcal{P}_{\mathcal{I}\mathcal{I}}^{\rm iso}(k_0)\,\delta(\ln k-\ln k_0)$, and the CMB/BAO physics is linear, PBH scenarios whose effective isocurvature spectrum, obtained from $\left<\mathcal{I}_{\rm DM}^2\right>(0)$  --- Eq.~\eqref{eq:IDM_final_result2} with $a=0$ multiplied by $\mathcal{S}^2(y_f)$ --- exceeds the bound in Fig.~\ref{Fig:power} are excluded.

In the $(M_{\rm PBH},f_{\rm PBH})$ plane, this yields the pink ({\it\bf CMB+BAO}) bounds in Fig.~\ref{Fig:constraints} for the two non-linear cut-offs.  These limits rule out SLABs as a significant DM component even for the more conservative non-linear cut-off (case~A).

The \textit{Planck} Collaboration~\cite{Planck:2018jri} also derived isocurvature constraints for three $k$ values
\begin{align}
\beta_{\rm iso}(k_{\rm CMB}{=}0.002\;\text{Mpc}^{-1}) &< 0.025, \notag\\
\beta_{\rm iso}(k_{\rm CMB}{=}0.05\;\text{Mpc}^{-1}) &< 0.26,\label{eq:beta_iso_constraints} \\
\beta_{\rm iso}(k_{\rm CMB}{=}0.1\;\text{Mpc}^{-1}) &< 0.47,\notag
\label{eq:beta_iso_3}
\end{align} 
where   $\beta_{\rm iso}(k) \equiv \mathcal{P}_{\mathcal{I}\mathcal{I}}^{\rm iso}(k) / [\mathcal{P}_{\mathcal{I}\mathcal{I}}^{\rm iso}(k) + \mathcal{P}_{\zeta}^{\rm adia}(k)]$ is the isocurvature fraction, with $\mathcal{P}_{\zeta}^{\rm adia}(k) \simeq 2.1\times 10^{-9} (k/k_\star)^{n_s-1}$, $n_s\simeq 0.965$, and $k_\star\simeq 0.05~{\rm Mpc}^{-1}$. Using only these three points (black dots in Fig.~\ref{Fig:power}) yields slightly weaker bounds (thin pink line in Fig.~\ref{Fig:constraints}), leading to the same conclusion that SLABs as a significant DM component are excluded.

\noindent{{\bf{\em Other PBH Constraints.}}}
The figure shows also other pre-existing constraints that can be divided into the following categories: for light PBHs ($\lesssim 10^{-16}\, M_\odot$), PBH evaporation via Hawking radiation is constrained by the gamma-ray background, Big Bang nucleosynthesis, CMB spectral distortions and Lyman-$\alpha$ data (magenta) (see Refs.~\cite{Saha:2021pqf,Laha:2019ssq,Ray:2021mxu,Mittal:2021egv,Clark:2016nst,Laha:2020ivk,Berteaud:2022tws,DeRocco:2019fjq,Dasgupta:2019cae,Calza:2021czr,Boudaud:2018hqb,Saha:2024ies,Khan:2025kag}). At slightly heavier masses, a small parameter-space island is constrained by the capture of PBHs by stars in ultra-faint dwarf galaxies (constraints labeled SC, yellow)\,\cite{Esser:2025pnt}. In the intermediate range ($\sim~10^{-11}\, M_\odot$ to tens of solar masses), microlensing surveys limit the PBH abundance from the non-detection of lensing events (red) (see Refs.\,\cite{Niikura:2017zjd,Niikura:2019kqi,Mroz:2024wag,Mroz:2024wia}). Around stellar masses, LIGO/Virgo/KAGRA GW detections constrain PBHs by comparing the observed merger rate with PBH merger predictions\,\cite{Andres-Carcasona:2024wqk} (orange) (see also Refs.\,\cite{LIGOScientific:2019kan,Kavanagh:2018ggo,Wong:2020yig,Hutsi:2020sol,DeLuca:2021wjr,Franciolini:2022tfm,DeLuca:2020fpg}). 
For PBHs above $\sim 1\,M_\odot$, constraints are obtained due to the modification of the CMB through enhanced ionization and spectral distortions from baryonic accretion and X-ray binaries (blue) (see Refs.\,\cite{Serpico:2020ehh,Agius:2024ecw,Murgia:2019duy,Manshanden:2018tze,Afshordi:2003zb}). Finally, at higher masses ($\gtrsim 10^2\,M_\odot$), dynamical constraints arise from the disruption of wide binaries, the stability of globular clusters, and the survival of dwarf galaxies (green) (see Refs.\,\cite{Carr:2018rid,Carr:2020erq}). The collapse of PBH-induced fluctuations can form too many clouds, dwarf galaxies and galaxy clusters (brown) (see~\cite{Carr:2018rid}), which would enhance the Ultraviolet
Luminosity Function (UV-LF) in tension with Hubble Space Telescope (purple) ~\cite{Gouttenoire:2023nzr} (see also ~\cite{Ellis:2025xju}), and would affect Lyman-$\alpha$ forests (Ly-$\alpha$)~\cite{Ivanov:2025pbu} (dark gold) (see also~\cite{Afshordi:2003zb,Murgia:2019duy}). We refer to \cite{Papanikolaou:2023cku} for constraints using magnetic field produced by SLAB accretion disks. The small green triangle in the top right corner is excluded from the CMB dipole anisotropy \cite{Carr:2020gox}. The gray region in bottom right corner shows the incredulity limit\,\cite{Carr:2020erq}, which corresponds to having fewer than one single PBH within the current Hubble volume.

\noindent{{\bf{\em Conclusions.}}} 
In this work, we have shown that the production of \textit{stupendously large} PBHs would generate large DM isocurvature perturbations, tightly constrained by CMB and BAO data. Using the upper bounds on the isocurvature power spectrum in Fig.~\ref{Fig:power}, we set the stringent limits shown in Fig.~\ref{Fig:constraints} in the mass range $M_{\rm PBH} \in [10^8,10^{19}]\,\msun$, excluding such objects as a significant DM component.

Our results strengthen the conclusion that the nano-Hertz gravitational-wave background detected in Pulsar Timing Arrays cannot be sourced by PBH mergers~\cite{Gouttenoire:2023nzr,Depta:2023uhy}. They also challenge scenarios attributing the early massive galaxy candidates observed by James Webb Space Telescope to PBH-induced fluctuations~\cite{Liu:2022bvr,Hutsi:2022fzw,Gouttenoire:2023nzr}.

Several extensions are possible. 
Our study assumes a monochromatic PBH mass function and could be generalized to extended mass distributions as in Ref.~\cite{Carr:2017jsz, Carr:2018rid}. Correlations between isocurvature and adiabatic modes, either of primordial origins or induced by cosmological evolution~\cite{Liu:2022okz}, could also be explored. In the case of full anticorrelation, limits on $\beta_{\rm iso}$ would drop to $\mathcal{O}(10^{-3})$~\cite{Planck:2018jri}, further tightening the abundance limits in the mass range considered. Next-generation CMB experiments, such as LiteBIRD~\cite{Matsumura:2013aja} and the Simons Observatory~\cite{SimonsObservatory:2018koc}, will improve the constraints presented here.
We leave for future work the extension of our constraints to the region $M_{\rm PBH}>10^{19}~M_\odot$ using the ringdown signal from the formation of such massive PBHs~\cite{DeLuca:2025uov,Yuan:2025bdp}, as well as searches for strong macrolensing in the CMB light.

\noindent{{\bf{\em Note added.}}}
Shortly before submission of this work, Ref.~\cite{Ivanov:2025pbu} closed the SLAB range $M_{\rm PBH}\in[10^{14},10^{16}]\,\msun$ using MIKE/HIRES Lyman-$\alpha$ data and EFT methods from Ref.~\cite{Ivanov:2023yla}, see region labeled {\it Ly-}$\alpha$ in Fig.~\ref{Fig:constraints}.

\noindent{{\bf{\em Acknowledgments.}}}
We thank B. Carr, P. Du, G. Perna, D. Perrone, A. Riotto, P. Schwaller, S. Trifinopoulos and V. Vaskonen for useful discussions. 
\\C.G., Y.G. and N.L. acknowledge support by the Cluster of Excellence ``PRISMA+'' funded by the German Research Foundation (DFG) within the German Excellence Strategy (Project No. 390831469). N.L. is grateful to the German Academic Scholarship Foundation for the award of a PhD fellowship.

\appendix
\onecolumngrid

\fontsize{11}{13}\selectfont

\makeatletter
\long\def\@makecaption#1#2{%
  \vskip\abovecaptionskip
  \parbox{\hsize}{%
    \fontsize{10.5}{12.5}\selectfont 
    \justifying
    #1.\ #2%
  }%
  \vskip\belowcaptionskip
}
\makeatother

\newpage
\vspace{1cm}

\ifarxiv
\begin{center}
\textbf{\it\Large Supplemental Material}
\end{center}
\renewcommand{\tocname}{\it
\vspace{-1.5cm}
}

\else
\begin{center}
\textbf{\large Supplemental Material for ``Closing the Mass Window for Stupendously Large Black Holes''}
\end{center}

\noindent
\renewcommand{\tocname}{\it
\vspace{-1cm}
Christopher Gerlach, Yann Gouttenoire, Antonio J. Iovino, Nicholas Leister
}
\fi

\titleformat{\section}
{\normalfont\fontsize{12}{14}\bfseries  \centering }{\thesection.}{1em}{}
\titleformat{\subsection}
{\normalfont\fontsize{12}{14}\bfseries \centering}{\thesubsection.}{1em}{}
\titleformat{\subsubsection}
{\normalfont\fontsize{12}{14}\bfseries \centering}{\thesubsubsection)}{1em}{}

\titleformat{\paragraph}
{\normalfont\fontsize{12}{14}\bfseries  }{\thesection:}{1em}{}

 {
 \hypersetup{linkcolor=black}
 \tableofcontents
 }
\section{Isocurvature Power Spectrum for extended PBHs}

\noindent In the main text, we have derived the isocurvature power spectrum  $\left<\mathcal{I}_{\rm DM}^2\right>$ associated with a population of PBHs,  assuming PBHs to be point-like. In this appendix, we derive $\left<\mathcal{I}_{\rm DM}^2\right>$, consistently accounting for their finite Schwarzschild horizon and spatial exclusion effects. This justifies the presence of the UV cut-off $\Theta\!\left(k_S(a)-k\right)$ in final formula for $\left<\mathcal{I}_{\rm DM}^2\right>$ in the main text.

\noindent For poisson-distributed objects with finite size, the PBH density contrast at formation can be expressed as
\begin{equation}
\label{eq:deltaPBH_exp}
   \delta_{\rm iso}^f(\mathbf{x})  = \Theta_{\rm PBH}(\mathbf{x}) \delta_{\rm S}-1,
\end{equation}
where $\delta_{\rm S}\equiv \rho_{\rm S}/\rho_{\rm PBH} = (n_{\rm PBH}V_S)^{-1}$ is
the ratio of the average energy density inside a PBH to the average energy density of the PBH gas with $n_{\rm PBH}$ the PBH number density, $\rho_{\rm S}\equiv M_{\rm PBH}/V_S$, $V_S\equiv 4\pi R_S^3/3$, and $R_S\equiv 2GM_{\rm PBH}\simeq 10^{-19}(M_{\rm PBH}/\msun)$ Mpc the density, volume and radius of a single PBH.  The function $\Theta_{\rm PBH}$ equals $1$ if $\mathbf{x}$ is contained inside a PBH and $0$ otherwise. We can rewrite $\Theta_{\rm PBH}$ as a sum over PBH centers,
\begin{equation}
\Theta_{\rm PBH}(\mathbf{x})=\sum_i \theta(\mathbf{x}-\mathbf{x}_i),
\qquad
\theta(\mathbf{x})=\Theta(R_S-|\mathbf{x}|).
\end{equation}
Its statistical average follows directly from homogeneity,
\begin{equation}
\left\langle \Theta_{\rm PBH}(\mathbf{x}) \right\rangle
=
n_{\rm PBH}\!\int d^3\mathbf{x}_0\,\theta(\mathbf{x}-\mathbf{x}_0)=
n_{\rm PBH}V_S,
\end{equation}
which implies $\langle\delta_{\rm iso}^f\rangle=0$ for $\delta_S=(n_{\rm PBH}V_S)^{-1}$. We model PBHs as randomly distributed hard spheres of radius $R_S$. The two-point function decomposes into contributions from the same PBH and from distinct PBHs,
\begin{equation}
\left\langle
\Theta_{\rm PBH}(\mathbf{x}_1)\Theta_{\rm PBH}(\mathbf{x}_2)
\right\rangle
=
n_{\rm PBH}V_\cap(r)
+
n_{\rm PBH}^2
\int d^3\mathbf{x}\,d^3\mathbf{y}\;
\theta(\mathbf{x}_1-\mathbf{x})\theta(\mathbf{x}_2-\mathbf{y})\,g(|\mathbf{x}-\mathbf{y}|),
\end{equation}
where $r\equiv|\mathbf{x}_1-\mathbf{x}_2|$, $V_\cap(r)$ is the overlap volume of two spheres of radius $R_S$, and $g(s)=1-\Theta(2R_S-s)$ enforces the hard-sphere exclusion. This yields
\begin{equation}
\label{eq:Theta_2pt_hardsphere}
\left\langle
\Theta_{\rm PBH}(\mathbf{x}_1)\Theta_{\rm PBH}(\mathbf{x}_2)
\right\rangle
=
n_{\rm PBH}^2
\Big[
V_S^2
+
n_{\rm PBH}^{-1}V_\cap(r)
-
V_{\rm excl}(r)
\Big],
\end{equation}
with $V_{\rm excl}(r)\equiv
\int_{|\mathbf{s}|<2R_S} d^3\mathbf{s}\;
V_\cap(|\mathbf{r}-\mathbf{s}|)$. Using Eq.~\eqref{eq:deltaPBH_exp}, one finds
\begin{equation}
\left\langle \delta_{\rm iso}^f(0)\delta_{\rm iso}^f(\mathbf{r})\right\rangle
=
\frac{V_\cap(r)}{n_{\rm PBH}V_S^2}
-\frac{V_{\rm excl}(r)}{V_S^2},
\end{equation}
which would be modified in the presence of initial clustering by the inclusion of the reduced PBH correlation function $\xi$\,\cite{Desjacques:2018wuu,Raidal:2017mfl,Ali-Haimoud:2018dau,MoradinezhadDizgah:2019wjf,Matsubara:2019qzv,Crescimbeni:2025ywm}. With the convention $\tilde X(k)\equiv\int d^3\mathbf r\,e^{i\mathbf k\cdot\mathbf r}X(\mathbf r)$, the overlap volume can be written as the convolution $V_\cap=\Theta_{R_S}*\Theta_{R_S}$ where $\Theta_{R_S}(r)\equiv \Theta(R_S-r)$, implying 
\[
\tilde{V}_\cap(k)=V_S^2W^2(kR_S),
\]
with $V_S\equiv 4\pi R_S^3/3$ and  $W(x)\equiv 3(\sin x-x\cos x)/x^3$. Since $V_{\rm excl}=\Theta_{2R_S}*V_\cap$ where $\Theta_{2R_S}(r)\equiv \Theta(2R_S-r)$, its transform factorizes as
\begin{equation}
\tilde{V}_{\rm excl}(r)
= \tilde\Theta_{2R}(k)\tilde{V_\cap}(k)
= \big[V_{2R}W(2kR_S)\big]\big[V_S^2W^2(kR_S)\big]
\end{equation}
with $V_{2R}\equiv 4\pi(2R_S)^3/3=8V_S$. Taking the Fourier transform of the above equation we get the power spectrum in Eq.~\eqref{eq:P_II_def}, which reads
\begin{equation}
\label{eq:delta_PBH_powerspectrum}
 \left<(\delta_{\rm iso}^{f})^2\right>
=
\frac{2}{3\pi}\left( \frac{k}{k_{\rm PBH}}\right)^3\mathcal{F}(k/k_S(a)),
\end{equation}
where $\mathcal{F}(x)\equiv W^2\!\left(x\right)\!\left[1-8\delta_S^{-1}W\!\left(2x\right)\right]$ with $W(x)\equiv 3(\sin x-x\cos x)/x^3$ is a cut-off function that verifies $\mathcal{F}(0)=1-8\delta_S^{-1}$ and $\mathcal{F}(x)\ll 1 $ for $x\gg 1$. In the regime relevant for the constraints, $\delta_S \gg 1$. Indeed, using $\gamma_H\equiv M_{\rm PBH}/M_f=\sqrt{\rho_f/\rho_S}$ and $\beta_f=(a_f/a_{\rm eq})f_{\rm PBH}\Omega_{\rm DM}/\Omega_{\rm m}<1$, one can express
\[
\delta_S(a_{\rm rec})=\frac{1}{\gamma_H^2\beta_f}\left(\frac{a_{\rm rec}}{a_f}\right)^3 ,
\]
which is much larger than unity since $\gamma_H\le1$, $\beta_f<1$, and PBHs form before recombination ($a_f<a_{\rm rec}$). The wavenumber $k$ is defined as comoving and $k_{\rm PBH}$ is the inverse comoving PBH interspacing\,\cite{Papanikolaou:2020qtd},  
\begin{equation}
\label{eq:kUV2}
k_{\rm PBH} = \left(\frac{4\pi }{3}n_{\rm PBH}\right)^{1/3} = \left(\frac{4\pi f_{\rm PBH} \,\rho_{\rm DM,0}}{3M_{\rm PBH}}\right)^{1/3},
\end{equation}
with $\rho_{\rm DM,0}\simeq 3\times10^{10} \,\msun\, \textrm{Mpc}^{-3}$ denoting the present-day mean dark matter density\,\cite{ParticleDataGroup:2020ssz}. The function $\mathcal{F}(k/k_S(a))$ implements a cutoff for $k\gtrsim k_S(a) \equiv R_S^{-1}a$ ---  assuming $a=1$ today --- due to PBH exclusion effects.  
Unlike usual physical scales, e.g. $k/a$ or $k_{\rm PBH}/a$, which redshift as $\propto a^{-1}$, the physical PBH radius $R_S$ is constant, implying that its comoving inverse scales as $k_S\propto a$ and is thus time-dependent. For a conservative estimate, we evaluate $k_S(a)$ at PBH formation, $a=a_f$, which is related to the comoving horizon scale at PBH formation, $k_f$, as  
\begin{equation}\label{eq:kf}
      k_f\,=\,\,\gamma_Hk_S(a_f)   \,\,\simeq\,\, 0.16~{\rm Mpc^{-1}}\gamma_{\rm H}^{\frac{1}{2}}\!\left(\frac{3.38}{g_\star(T_f)}\!\right)^{\!\frac{1}{12}}\!\!\left(\!\frac{10^{15}M_\odot}{M_{\rm PBH}}\!\right)^{\!\frac{1}{2}}\!\!\!,
\end{equation}
with $\gamma_H$ and $g_{\star}(k)$ that denote, respectively, the PBH mass in unit of the horizon mass (we fix $\gamma_{\rm H} = 1$ for definiteness) and the number of effective relativistic degrees of freedom, both evaluated at PBH formation. For simplicity, in the main text, we approximate the function $\mathcal{F}(k/k_S(a))$ in Eq.~\eqref{eq:delta_PBH_powerspectrum} by an Heaviside distribution $\Theta\!\left(k_S(a)-k\right)$.

\section{Evolution of $\Phi$ in a radiation--DM background}
\label{app:Transfer_function}

\noindent 
In this appendix, we derive --- following Ref.~\cite{Peter:2013avv} --- the coupled evolution of the Newtonian potential and DM isocurvature mode~\footnote{In the case of interest, DM is composed of a fraction $f_{\rm PBH}$ of PBHs and a fraction $1 - f_{\rm PBH}$ of particle DM.}.

\medskip

\noindent We work in Newtonian gauge,
\begin{equation}
ds^2=a^2(\eta)\Big[(1+2\Phi)d\eta^2-(1-2\Phi)\delta_{ij}dx^i dx^j\Big],
\end{equation}
assuming vanishing anisotropic stress. The background evolution is governed by
\begin{equation}
\mathcal H^2=\frac{8\pi G}{3}a^2\bar\rho,
\qquad
\mathcal H'=-\frac{1+3\omega}{2}\mathcal H^2,
\end{equation}
where $\mathcal H\equiv a'/a$ and $\omega\equiv \bar p/\bar\rho$. At linear order, Einstein equations yield
\begin{align}
\nabla^2 \Phi - 3\mathcal H(\Phi' + \mathcal H \Phi) &= \frac{3}{2}\mathcal H^2\,\delta, \\
\Phi'' + 3\mathcal H \Phi' -3\omega \mathcal H^2 \Phi &= \frac{3}{2}\mathcal H^2\,\frac{\delta p}{\bar\rho},
\end{align}
with $\delta\equiv \delta\rho/\bar\rho$. Introducing
\begin{equation}
c_s^2\equiv \frac{\bar p'}{\bar\rho'},
\qquad
\delta_{p,\rm nad}\equiv \frac{\delta p-c_s^2\delta\rho}{\bar\rho},
\end{equation}
one obtains
\begin{equation}
\label{eq:Phi_master_short}
\Phi''+3\mathcal H(1+c_s^2)\Phi'
+\Big[3(c_s^2-\omega)\mathcal H^2-c_s^2\Delta\Big]\Phi
=
\frac{3}{2}\mathcal H^2\,\delta_{p,\rm nad}.
\end{equation}

\medskip

\noindent We now consider $\bar\rho=\bar\rho_r+\bar\rho_{\rm DM}$ with
\begin{equation}
\omega_r=c_r^2=\tfrac13,
\qquad
\omega_{\rm DM}=c_{\rm DM}^2=0.
\end{equation}
Using $\bar\rho_r \propto a^{-4}$ and $\bar\rho_{\rm DM} \propto a^{-3}$, we define
\begin{equation}
y \equiv \frac{a}{a_{\rm eq}},
\end{equation}
where $a_{\rm eq}$ denotes the scale factor at matter–radiation equality. This yields
\begin{equation}
\Omega_{\rm DM} = \frac{y}{1+y}, \quad
\Omega_r = \frac{1}{1+y}, \quad
\omega = \frac{1}{3(1+y)}, \quad
c_s^2 = \frac{1}{3\left(1+\frac{3}{4}y\right)}.
\end{equation}
\noindent The DM--radiation isocurvature mode is
\begin{equation}
\mathcal{I}_{\rm DM}\equiv \delta_{\rm DM}-\frac{3}{4}\delta_r,
\end{equation}
and the associated non-adiabatic pressure perturbation reads
\begin{equation}
\delta_{p,\rm nad}
=
-\frac{y}{(1+y)(4+3y)}\,\mathcal{I}_{\rm DM}.
\end{equation}
Equation~\eqref{eq:Phi_master_short} therefore becomes
\begin{equation}
\label{eq:Phi_short}
\Phi''+3\mathcal H(1+c_s^2)\Phi'
+\Big[3(c_s^2-\omega)\mathcal H^2-c_s^2\Delta\Big]\Phi
=
-\frac{3}{2}\mathcal H^2\frac{y}{(1+y)(4+3y)}\mathcal{I}_{\rm DM}.
\end{equation}

\medskip

\noindent Combining the fluid equations for radiation and DM yields
\begin{equation}
\label{eq:S_short}
\mathcal{I}_{\rm DM}''+\mathcal H \mathcal{I}_{\rm DM}'-\frac{y}{4+3y}\Delta \mathcal{I}_{\rm DM}
=
-\frac{8}{3(4+3y)\mathcal H^2}\Delta^2\Phi.
\end{equation}

\medskip
\noindent The background evolution reads
\begin{equation}
\mathcal H=k_{\rm eq}\sqrt{\frac{1+y}{2y^2}},
\qquad
y(\eta)=\left(\frac{\eta}{\eta_\star}\right)^2+2\frac{\eta}{\eta_\star},
\qquad
\eta_\star=\frac{2\sqrt2}{k_{\rm eq}},
\end{equation}
where $k_{\rm eq}\equiv \mathcal H_{\rm eq}=\sqrt{2z_{\rm eq}\Omega_m}H_0\simeq (98~\rm Mpc)^{-1}$, with $z_{\rm eq}\simeq 3402$, $\Omega_m\simeq 0.315$, and $H_0 \simeq 67.4~{\rm km/s/Mpc}$~\cite{ParticleDataGroup:2020ssz}. 
Introducing $\kappa\equiv k/k_{\rm eq}$ and taking $y$ as the time variable, derivatives transform as
\begin{equation}
X' = y\mathcal H\,\dot X,
\qquad
X'' = y^2\mathcal H^2\,\ddot X + \frac{1-3\omega}{2}y\mathcal H^2\,\dot X.
\end{equation}
In terms of $y$, Eqs.~\eqref{eq:Phi_short} and \eqref{eq:S_short} become
\begin{align}
\label{eq:Phi_eq}
\ddot{\Phi}_{\mathbf k}
+\frac{1}{2y}\left(7-\frac{1}{1+y}+\frac{8}{4+3y}\right)\dot{\Phi}_{\mathbf k}
+\frac{3+8\kappa^2 y}{3y(1+y)(4+3y)}\Phi_{\mathbf k}
&=
-\frac{2}{y(1+y)(4+3y)}\mathcal{I}_{\rm DM,\mathbf k},
\\
\label{eq:S_eq}
\ddot{{\mathcal{I}}}_{\rm DM,\mathbf k}
+\frac{2+3y}{2y(1+y)}\dot{{\mathcal{I}}}_{\rm DM,\mathbf k}
+\frac{2y\kappa^2}{(1+y)(4+3y)}\mathcal{I}_{\rm DM,\mathbf k}
&=
-\frac{8y^2\kappa^4}{3(1+y)(4+3y)}\Phi_{\mathbf k}.
\end{align}

\noindent These equations govern the evolution from radiation domination ($y\ll1$) to DM domination ($y\gg1$), and apply to both adiabatic ($\mathcal{I}_{\rm DM}=0$) and isocurvature ($\mathcal{I}_{\rm DM}\neq0$) initial conditions.

\begin{figure}[t]
\centering
\includegraphics[width=0.48\textwidth]{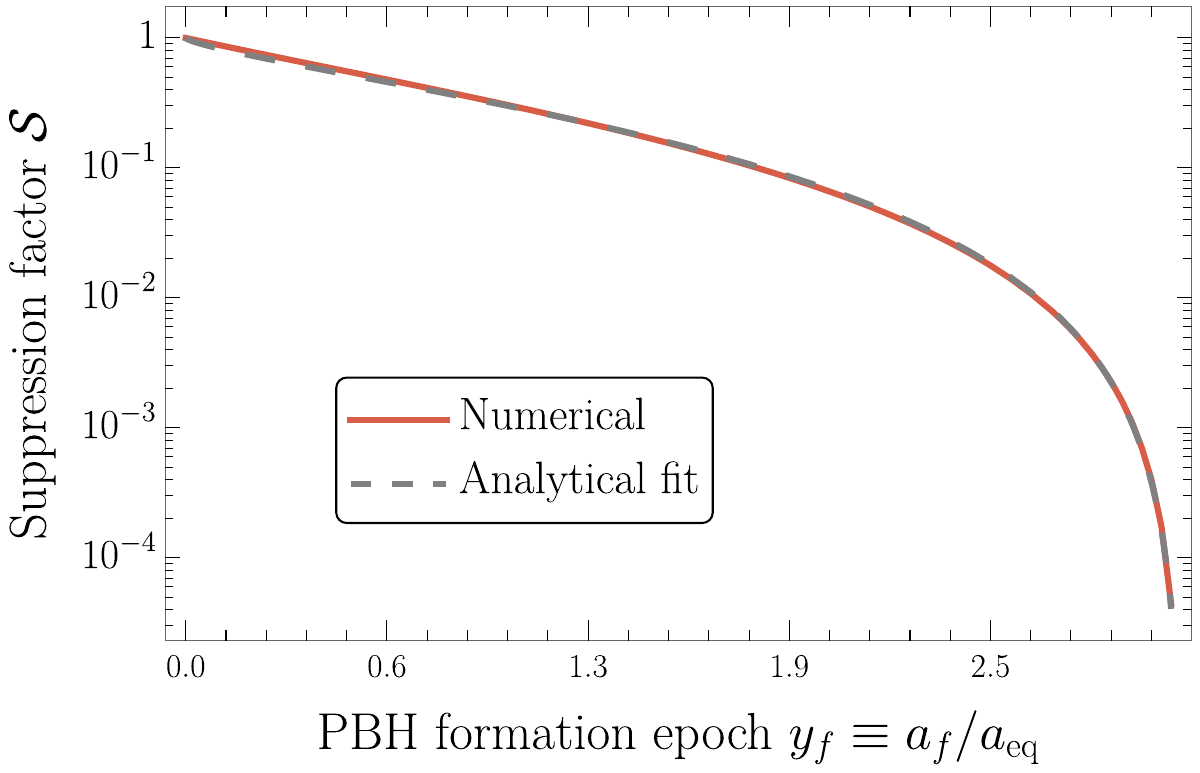}
\includegraphics[width=0.48\textwidth]{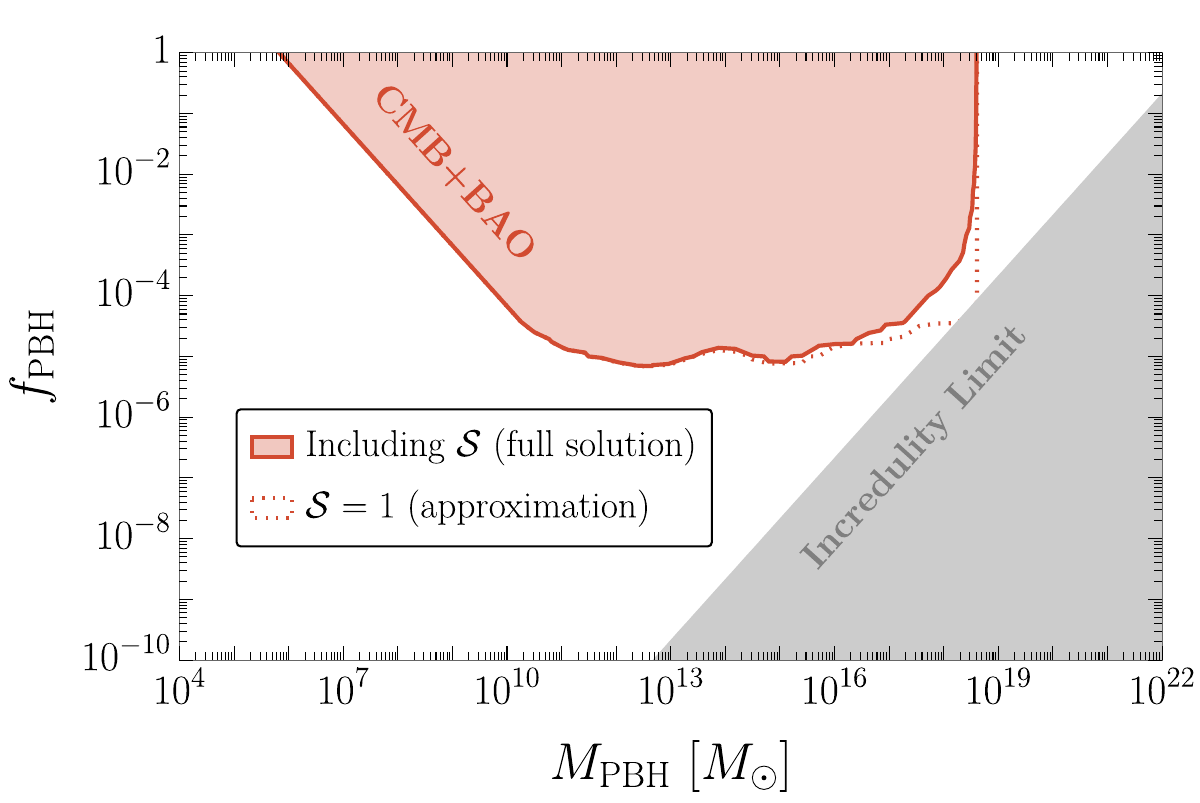}
\caption{\label{fig:supp_fac}\it
\textbf{Left:} Suppression factor $\mathcal{S}(y_f)$ defined in Eq.~\eqref{eq:supp_S}, encoding the effect of finite-time PBH formation on isocurvature constraints. The analytical fit in Eq.~\eqref{eq:supp_S_ana} accurately reproduces the numerical result. As expected, $\mathcal{S}\to 1$ for $y_f\to 0$, recovering the standard primordial case, while $\mathcal{S}\to 0$ for $y_f\to 1$, where isocurvature perturbations have no time to source curvature.
\textbf{Right:} Corresponding impact on the \textit{Planck} constraints obtained from Eq.~\eqref{eq:supp_S_Planck}. Finite-time generation of isocurvature modes leads to a systematic, though moderate, weakening of the bounds compared to the standard treatment with primordial initial conditions.}
\end{figure}

\section{Finite-time generation of isocurvature and suppression of Planck constraints}

\noindent In this appendix, we quantify how the finite-time generation of isocurvature perturbations suppresses their late-time amplitude and determine the corresponding impact on \textit{Planck} constraints.
 
\noindent The isocurvature bounds quoted in Ref.~\cite{Buckley:2025zgh} are obtained by evolving primordial initial conditions with a Boltzmann solver, where the isocurvature mode is specified deep in the radiation era at $y=0$. In our setup, instead, isocurvature perturbations are generated at PBH formation at $y_f$ which may occur significantly later. This leads to a suppression of the dark-matter isocurvature power spectrum,
\begin{equation}
\label{eq:supp_S_Planck}
\langle \mathcal{I}_{\rm DM}^2 \rangle 
~\rightarrow~
\mathcal{S}^2(y_f)\,
\langle \mathcal{I}_{\rm DM}^2 \rangle .
\end{equation}

\noindent 
We define the suppression factor as
\begin{equation}
\label{eq:supp_S}
\mathcal{S}_{\mathbf{k}}(y_f)
\equiv
\frac{\mathcal{I}_{\rm DM,\mathbf{k}}(0)}{\mathcal{I}_{\rm DM,\mathbf{k}}(y_f)},
\end{equation}
and determine it by requiring that the Newtonian potential at recombination,
$y_{\rm rec}\equiv a_{\rm rec}/a_{\rm eq}\simeq 3.1$, which controls the observed CMB anisotropies, is unchanged. This fixes
\begin{equation}
\label{eq:S_num}
\Phi_{\mathbf{k}}(y_{\rm rec};\,\mathcal{I}_{\rm DM,\mathbf{k}}(0))
=
\Phi_{\mathbf{k}}(y_{\rm rec};\,\mathcal{I}_{\rm DM,\mathbf{k}}(y_f)),
\end{equation}
where $\Phi_{\mathbf{k}}(y;\,\mathcal{I}_{\rm DM,\mathbf{k}}(y_i))$ denotes the solution of
Eqs.~\eqref{eq:Phi_eq} and \eqref{eq:S_eq} with initial conditions
$\mathcal{I}_{\rm DM,\mathbf{k}}(y_i)$ and $\Phi_{\mathbf{k}}(y_i)=0$.
Equivalently, $\mathcal{S}_{\mathbf{k}}(y_f)$ rescales the isocurvature amplitude such that
a mode generated at $y_f$ produces the same late-time gravitational potential, $\Phi_{\mathbf{k}}(y_{\rm rec})$, as a mode
initialized deep in the radiation era at $y=0$.

\noindent We compute $\mathcal{S}_{\mathbf{k}}(y_f)$ by numerically solving Eqs.~\eqref{eq:Phi_eq} and \eqref{eq:S_eq}. As shown in Fig.~\ref{fig:supp_fac} (left), it is well approximated by
\begin{equation}
\label{eq:supp_S_ana}
\mathcal{S}(y_f) \simeq \left[1-\left(\frac{y_f}{y_{\rm rec}}\right)^{b/a}\right]^a,
\qquad
a\simeq 2.05\approx 2,\quad b\simeq 1.48\approx \frac{3}{2},
\end{equation}
where $y_{\rm rec}\simeq 3.1$ and the expression for $y_f$ is given in the main text.
We find the residual dependence on $\mathbf{k}$ to be negligible, and therefore omit the $\mathbf{k}$ subscript. 

\medskip

\noindent \textit{Planck} constrains DM isocurvature through its contribution to the Newtonian potential at recombination $\Phi(y_{\rm rec})$. In our scenario, isocurvature perturbations $\mathcal{I}_{\rm DM}$ are generated at a finite time $y_f$ and thus have less time to source metric perturbations. We account for this by mapping the spectrum at finite time, $\mathcal{I}_{\rm DM}(y_f)$, onto an effective primordial one, $\mathcal{I}_{\rm DM}(0)$, that reproduces the same $\Phi(y_{\rm rec})$. This mapping is encoded in $\mathcal{S}(y_f)$, which quantifies the enhancement of $\mathcal{I}_{\rm DM}(y_f)/\mathcal{I}_{\rm DM}(0)$ required to match a fixed late-time potential. Consequently, the upper limit on $\mathcal{I}_{\rm DM}(y_f)$ is relaxed with respect to the one on $\mathcal{I}_{\rm DM}(0)$, implying weaker bounds. Equivalently, the \textit{Planck} constraints are suppressed by $\mathcal{S}^2(y_f)$ as in Eq.~\eqref{eq:supp_S_Planck}. The resulting impact is shown in Fig.~\ref{fig:supp_fac} (right).

\bibliography{main}

\end{document}